\documentclass[showpacs,preprintnumbers,amsmath,amssymb,twocolumn,superscriptaddress,prb]{revtex4}
\usepackage{times}
\usepackage{graphicx}
\usepackage{amsfonts}
\usepackage{amsmath, amsthm, amssymb}
\usepackage{dsfont}
\usepackage{subfigure}

\newcommand{\avec}{\hat{\mathbf{a}}} 
\newcommand{\bvec}{\hat{\mathbf{b}}} 
\newcommand{\vvec}{\hat{\mathbf{v}}}
\newcommand{\uvec}{\hat{\mathbf{u}}} 

\newcommand{\vk}{\mathbf{k}} 
\newcommand{\vp}{\mathbf{p}} 
\newcommand{\vpf}{\mathbf{p}_\text{F}}
\newcommand{\vecr}{\mathbf{r}}

\newcommand{\e}[1]{\mathrm{e}^{#1}}

\newcommand{\cop}{\hat{c}} 

\newcommand{\ie}{i.e.}
\newcommand{\eg}{\textit{e.g. }}
\newcommand{\etal}{\emph{et al. }}
\def\i{\mathrm{i}}

\begin{document}
\title[Density of states near a vortex core in ferromagnetic superconductors: Application to STM measurements]{Density of 
states near a vortex core in ferromagnetic superconductors: Application to STM measurements}
\author{Jacob Linder}
\affiliation{Department of Physics, Norwegian University of
Science and Technology, N-7491 Trondheim, Norway}
\author{Takehito Yokoyama}
\affiliation{Department of Applied Physics, Nagoya University, Nagoya, 464-8603, Japan}
\author{Asle Sudb{\o}}
\affiliation{Department of Physics, Norwegian University of
Science and Technology, N-7491 Trondheim, Norway}

\date{Received \today}
\begin{abstract}
We investigate numerically the local density of states (LDOS) in the vicinity of a vortex core in a ferromagnetic superconductor. 
Specifically, we investigate how the LDOS is affected by the relative weight of the spin bands in terms of the superconducting 
pairing, and we also examine the effect of different pairing symmetries for the superconducting order parameter. Our findings 
are directly related to scanning tunneling microscopy measurements and may thus be highly useful to clarify details of the 
superconducting pairing in recently discovered ferromagnetic superconductors.
\end{abstract}

\maketitle

\section{Introduction}

Recently, UCoGe was added to the distinguished list of materials (already featuring UGe$_2$ and URhGe) which appear to 
display coexistence of ferromagnetism and superconductivity \cite{huy,saxena,aoki}. While ferromagnetism and conventional 
superconductivity may be shown to be antagonistic in terms of a bulk coexistent state \cite{breakdown}, several studies 
have pointed out the possibility of a non-unitary, spin-triplet superconducting state coexisting with itinerant ferromagnetism 
\cite{walker,harada,nevidomskyy,linder1,linder2,shopova2005}. The synthesis of two important phenomena in condensed-matter 
physics, ferromagnetism and superconductivity, is not only interesting from the point of view of basic research, but 
has also spawned hope of potential applications in low-temperature nanotechnology.
\par
A number of questions arise concerning the nature of the coexistence of ferromagnetic and superconducting order. In particular, it 
is crucial to address \textit{i)} whether the two long-range orders are phase-separated or not,  \textit{ii)} whether the microscopic 
coexistence is spatially homogeneous or not, and \textit{iii)} what the symmetry of the superconducting order parameter is. Concerning 
the first question, the answer clearly appears to be 'yes', since the onset of superconductivity appears inside the ferromagnetic part 
of the phase diagram \cite{harada}. The second question is, however, still open. Some authors have studied spatially uniform coexistence 
of ferromagnetic and superconducting order \cite{nevidomskyy,linder1,linder2,shopova2005,uzunov}, while others have pointed out the 
intriguing possibility of a spontaneously formed vortex lattice state\cite{tewari2004, leo, knigavko}, due to the internal field. 
It has been argued \cite{mineev} that a key factor with regard to whether such a spontaneous vortex phase appears or not is the 
magnitude of the internal magnetization $\mathbf{M}$. Finally, although the issue of pairing symmetry raised in the third question 
has not been established conclusively, the most likely option appears to be a non-unitary, spin triplet superconducting state, 
where the spin of the Cooper pair couples to the bulk magnetization through a third order term 
$\sim \i(\mathbf{d}_\vk\times\mathbf{d}_\vk^*)\cdot \mathbf{M}$ in the Ginzburg-Landau free energy. Several 
studies \cite{brataas,shen,gronsleth,linder3,yokoyama} have addressed means by which one may identify the 
pairing symmetry of the superconducting order parameter in a ferromagnetic superconductor, mainly focusing 
on transport properties.
\par
Clearly, it would be highly desirable to clarify experimental signatures of a possible spontaneous vortex lattice-phase realized 
in a ferromagnetic superconductor. In this work, we present numerical results for the local density of states (LDOS) in the vicinity of 
a vortex-core of a ferromagnetic superconductor. Our approach is based on the quasiclassical theory of superconductivity, and 
takes into account several crucial factors such as the depletion of the order parameter near the vortex core in addition to 
self-consistently obtained magnetic and superconducting order parameters. Our results are directly relevant for scanning 
tunneling microscopy (STM) measurements,\cite{Hess89} and may be useful to clarify signatures of the existence of a 
spontaneously formed vortex lattice and also the pairing symmetry of the superconducting order parameter.

\begin{figure}[h!]
\centering
\resizebox{0.42\textwidth}{!}{
\includegraphics{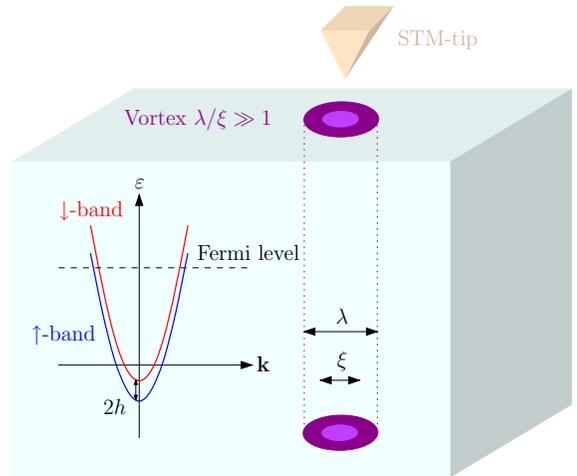}}
\caption{(Color online) Schematic illustration of the model.}
\label{fig:ratio}
\end{figure}

\par
This paper is organized as follows. In Sec. \ref{sec:theory}, we establish the theoretical framework employed in 
this work. Namely, we use the quasiclassical approximation and solve the Eilenberger equation in the vicinity of 
the vortex core with appropriate boundary conditions. In Sec. \ref{sec:results}, we present our results for the 
spatial- and energy-dependence of the local density of states near the vortex core. Specifically, we investigate 
how the relative weight of the spin bands in terms of the superconducting pairing and different pairing symmetries 
for the superconducting order parameter affect the density of states. In Sec. \ref{sec:discuss} and \ref{sec:summary}, 
we discuss and summarize the main results of the paper. We will use boldface notation for 2-vectors, $\hat{\ldots}$ for $4\times4$ matrices, and $\underline{\ldots}$ for $2\times2$ matrices.

\section{Theoretical framework}\label{sec:theory}

It is generally believed that the pairing symmetry in ferromagnetic superconductors may be classified as a non-unitary, 
spin-triplet state. \cite{walker, nevidomskyy,linder1} Our starting point is the quasiclassical Eilenberger equation 
\cite{eilenberger} for such a system, which in the clean limit reads (see Appendix \ref{app:eilenberger} for details)
\begin{align}\label{eq:eilenberger}
\i\mathbf{v}_F\cdot\nabla \hat{g}^R + [\varepsilon\hat{\rho}_3 +\hat{M} + \hat{\Delta}(\vpf), \hat{g}^R]=0,
\end{align}
where $\varepsilon$ is the quasiparticle energy measured from the Fermi level, $\mathbf{v}_F$ is the Fermi 
velocity, and $[\ldots]$ is a commutator. The exchange field $h$ and the superconducting order 
parameters $\Delta_\sigma$ are contained in the terms of $\hat{M} = h\text{diag}(\underline{\tau_3},\underline{\tau_3})$ 
in addition to
\begin{align}\label{eq:ttt}
\hat{\Delta}(\vp_F) &= \begin{pmatrix} 
\underline{0} & \underline{\Delta}(\vp_F)\\
-\underline{\Delta}^*(\vp_F) & \underline{0}\\
\end{pmatrix},\notag\\
\underline{\Delta}(\vp_F) &= \begin{pmatrix} 
\Delta_\uparrow(\vp_F) & 0\\
0 & \Delta_\downarrow(\vp_F) \\
\end{pmatrix}.
\end{align}
The matrices $\hat{\rho}_i$ and $\underline{\tau_i}$ are defined in the Appendix.
The retarded part of the Green's function, $\hat{g}^R$, will have the structure
\begin{align}
\hat{g}^R = \begin{pmatrix}
\underline{g}(\mathbf{r},\vp_F,\varepsilon) & \underline{f}(\mathbf{r},\vp_F,\varepsilon) \notag\\
-\underline{f}^*(\mathbf{r},-\vp_F,-\varepsilon) & -\underline{g}^*(\mathbf{r},-\vp_F,-\varepsilon)\notag\\
\end{pmatrix},
\end{align}
and must satisfy the normalization condition $(\hat{g}^R)^2=\hat{1}$. Due to the internal symmetry relations 
between the components of $\hat{g}^R$, one may parametrize it very conveniently by means of a so-called 
Ricatti-parametrization \cite{schopohl,eschrigprb}. In the absence of interband-scattering, the Eilenberger equation decouples into 
two $2\times2$ equations as follows:
\begin{align}\label{eq:eilen2}
\i\mathbf{v}_F\cdot\nabla\underline{g_\sigma} + [\varepsilon \underline{\tau_3} + \sigma h\underline{\tau_0} + \underline{\Delta_\sigma}(\vp_F), \underline{g_\sigma}] = 0,
\end{align}
where we have introduced
\begin{align}\label{eq:tt}
\underline{g_\sigma} &= N_\sigma\begin{pmatrix}
1 - a_\sigma b_\sigma & 2a_\sigma \\
2b_\sigma & - 1 + a_\sigma b_\sigma\\
\end{pmatrix},\; N_\sigma = (1+a_\sigma b_\sigma)^{-1},\notag\\
&\underline{\Delta_\sigma}(\vp_F) = \begin{pmatrix}
0 & \Delta_\sigma(\vp_F)\\
-\Delta_\sigma^*(\vp_F) & 0 \\
\end{pmatrix}.
\end{align}
Note that the gap matrix in Eq. (\ref{eq:tt}) is a $2\times2$ matrix in particle-hole space, while the gap matrix in Eq. (\ref{eq:ttt}) is a $2\times2$ matrix in spin-space. From Eq. (\ref{eq:eilen2}), one obtains two decoupled differential equations for $a_\sigma$ and $b_\sigma$: 
\begin{align}
\i \mathbf{v}_F\cdot\nabla a_\sigma &+2a_\sigma\varepsilon - a_\sigma^2\Delta_\sigma^*(\vp_F) - \Delta_\sigma(\vp_F)=0,\notag\\
\i \mathbf{v}_F\cdot\nabla b_\sigma &-2b_\sigma\varepsilon - b_\sigma^2\Delta_\sigma(\vp_F) - \Delta_\sigma^*(\vp_F)=0.
\end{align}
Note that the above equations do not have any \textit{explicit} dependence on the exchange splitting $h$. As we shall 
see later, the exchange splitting does however enter implicitly through the spin-dependent gaps $\Delta_\sigma$. Note 
that the magnetic vector potential $\mathbf{A}$ may be incorporated above simply by a shift in the quasiparticle 
energies: $\varepsilon\to \varepsilon + e\mathbf{v}_F\cdot\mathbf{A}$. In a gauge that renders the superconducting 
gaps to be real, one finds that $e\mathbf{A}\to e\mathbf{A} - \nabla\Phi/2$, where $\Phi$ is the superconducting 
phase associated with the broken U(1) symmetry. Therefore, the total Doppler shift in the quasiparticle energies 
is $\varepsilon \to \varepsilon - em\mathbf{v}_F\cdot\mathbf{v}_s$, where the gauge-invariant superfluid velocity 
is $\mathbf{v}_s = (\nabla\Phi - 2e\mathbf{A})/(2m)$. Below, 
we keep the distribution of the superconducting phase in the order parameter and consider the case with 
Ginzburg-Landau parameter 
$\kappa \gg 1$, for which the magnetic vector potential $\mathbf{A}$ may be neglected. This follows since
we are considering only one single vortex, i.e. the zero-field limit, such that only gauge-field fluctuations
around zero could possibly be relevant. However, assuming that the superconductors are strongly type-II with 
$\kappa \gg 1$, gauge-field fluctuations are suppressed \cite{zbt1999,nguyen-sudbo1999}.  
\par

\begin{figure}[h!]
\centering
\resizebox{0.48\textwidth}{!}{
\includegraphics{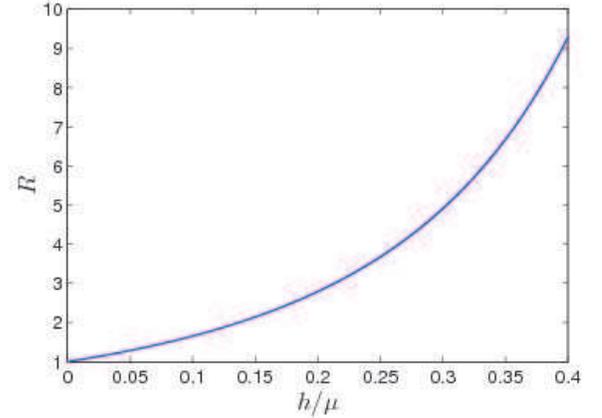}}
\caption{(Color online) The ratio $R$ between the majority- and minority-spin gaps as a function of $h/\mu$ 
as obtained from a self-consistent, mean-field solution [Eq. (\ref{eq:hmu})].}
\label{fig:ratio}
\end{figure}
\par
In order to solve the above Ricatti-equations, we follow closely the procedure of Ref.~\onlinecite{schopohl}. 
Let us consider the term with $\mathbf{v}_F\cdot\nabla$ in more detail. Assume that we have a cylindrically 
symmetric vortex situated at $r_a=r_b=0$ with its axis along $\hat{\mathbf{c}}$. The position vector in this 
coordinate system then reads
$\mathbf{r} = r_a\avec + r_b\bvec$. Assuming that the transport of quasiparticles primarily takes place in 
the $\avec-\bvec$-plane, we may define the Fermi velocity as 
\begin{align}
\mathbf{v}_F = v_F(\cos\theta\avec + \sin\theta\bvec) \equiv v_F\vvec
\end{align}
and its orthogonal vector $\uvec = -\sin\theta\avec + \cos\theta\bvec$. Thus, the position vector $\mathbf{r}$ 
may also be expressed as $\mathbf{r} = x\vvec + y\uvec$, where we have defined 
\begin{align}
x=r_a\cos\theta + r_b\sin\theta,\; y=-r_a\sin\theta + r_b\cos\theta.
\end{align}
Using the new coordinate system $\vvec-\uvec$, the Ricatti equations may be rewritten as
\begin{align}
\i v_F\partial_x a_\sigma &+ [2\varepsilon - \Delta_\sigma^*a_\sigma]a_\sigma - \Delta_\sigma = 0,\notag\\
\i v_F\partial_x b_\sigma &- [2\varepsilon + \Delta_\sigma b_\sigma]b_\sigma - \Delta_\sigma^* = 0,
\end{align}
where $a_\sigma = a_\sigma(x,y)$ and $\Delta_\sigma = \Delta_\sigma(x,y)$. The above equations may be solved by imposing boundary conditions for $\{a_\sigma,b_\sigma\}$ in the bulk of the superconductor. The Ricatti-equations with $\varepsilon>0$ for $a_\sigma$ and $b_\sigma$ are stable for integration from $x\to(-\infty)$ and $x\to\infty$, respectively (opposite for $\varepsilon<0$).\cite{schopohl} The boundary conditions then read: 
\begin{align}
a_\sigma[x\to(-\infty)] &= (\varepsilon - \sqrt{\varepsilon^2-|\Delta_\sigma|^2})/\Delta_\sigma^*,\notag\\
b_\sigma(x\to\infty) &= -(\varepsilon - \sqrt{\varepsilon^2-|\Delta_\sigma|^2})/\Delta_\sigma.
\end{align}
The superconducting order parameter $\Delta_\sigma$ is now modelled in the presence of a vortex centered at $r_a=r_b=0$. 
In general, the superconducting order parameter may be written as\cite{yokoyama2}
\begin{align}
\Delta_\sigma(\vecr,\theta,\varepsilon) &= \Delta_{\sigma,0}\chi_\sigma(\theta,\varepsilon) F(r) \e{\i m\phi},
\end{align}
assuming a vorticity $m$. Here, $\Delta_0$ is the gap magnitude, $\chi(\theta,\varepsilon)$ is a symmetry factor for the 
gap (taking into account both anisotropicity and frequency-dependence), $F(r)$ models the spatial depletion of the gap 
near the vortex core, while $\tan\phi = r_b/r_a$. We will here restrict our attention to an even-frequency, $p$-wave 
symmetry, which is believed to be the most likely candidate for the order parameter in ferromagnetic superconductors. 
Assuming that the angular symmetry is the same for both the majority and minority spin gaps and considering the usual 
case of $m=1$, we explicitly have
\begin{align}
\Delta_\sigma(\vecr,\theta) = \Delta_{\sigma,0}\chi(\theta) \tanh\Big(\frac{\sqrt{x^2+y^2}}{\xi}\Big) \frac{x+\i y}{\sqrt{x^2+y^2}}.
\end{align}
In what follows, we will compare the cases $\chi(\theta)=\cos\theta$ and $\chi(\theta) = \e{\i\theta}$, 
and also investigate how the LDOS changes depending the relative weight of the superconducting instability in both 
spin-bands. The normalized LDOS for spin species $\sigma$ is given by
\begin{align}
N_\sigma(\vecr,\varepsilon) = \int^{2\pi}_0 \frac{\text{d}\theta}{2\pi} \text{Re}\{ (1-a_\sigma b_\sigma)/(1+a_\sigma b_\sigma)\},
\end{align}
and we introduce the total LDOS in the standard way as 
\begin{align}
N(\vecr,\varepsilon) = \sum_\sigma N_\sigma(\vecr,\varepsilon)/2.
\end{align}
To account for a finite quasiparticle lifetime $\tau$, we let $\varepsilon\to\varepsilon+\i\delta$ where $\delta\sim \tau^{-1}$. 
From now on, we fix $\delta=0.1\Delta_{\uparrow,0}$ and comment further upon the role of inelastic scattering in Appendix \ref{app:inelastic}.
\par
Even if the exchange field $h$ is absent from the Eilenberger equation, the LDOS is \textit{not} independent of the value 
of $h$. The reason for this is that the magnitude of the superconducting gaps depend on the strength of the exchange 
splitting. Following the approach of Refs.~\onlinecite{linder1,linder2}, we derive from a weak-coupling mean-field 
theory that the self-consistent solution of bulk superconducting gaps in the $T\to 0$ limit may be written as \begin{align}\label{eq:gap}
\Delta_{\sigma,0}/\omega_0 = c\exp[-1/(g\sqrt{1+\sigma h/\mu})],
\end{align}
 where the prefactor is equal to $c\simeq 2.43$ for a $p_x$-wave symmetry $[\chi(\theta)=\cos\theta]$, $c = 2.00$
for a chiral $p$-wave symmetry $[\chi(\theta) = \e{\i\theta}]$.\cite{linder1,linder2} Here, $g=V_0N_0$ is the weak-coupling constant which we set to $g = 0.2$ and $\omega_0$ is the typical frequency width around Fermi level for the bosons responsible for the 
superconducting pairing. Above, $V_0$ is the strength of the pairing interaction, $N_0$ is the LDOS at Fermi level in the normal-state and $\mu$ denotes the Fermi energy. The reader may consult Appendix \ref{app:gap} for a derivation of Eq. (\ref{eq:gap}). We find that the ratio between the majority- and minority-spin gaps may be 
written as
\begin{align}\label{eq:hmu}
\frac{\Delta_{\uparrow,0}}{\Delta_{\downarrow,0}} \equiv R(h/\mu) = \exp\Bigg[\frac{\sqrt{1 + h/\mu} - \sqrt{1-h/\mu}}{g\sqrt{1-(h/\mu)^2}}\Bigg]
\end{align}
when assuming that $h/\mu \in [0,1)$ (shown in Fig. \ref{fig:ratio}). In UGe$_2$, the energy splitting between the majority 
and minority spin bands was estimated \cite{saxena} to lie around 70 meV, which yields $R\simeq 1.42$ when assuming $\mu = 1$ eV. 

\begin{figure}[h!]
\centering
\resizebox{0.48\textwidth}{!}{
\includegraphics{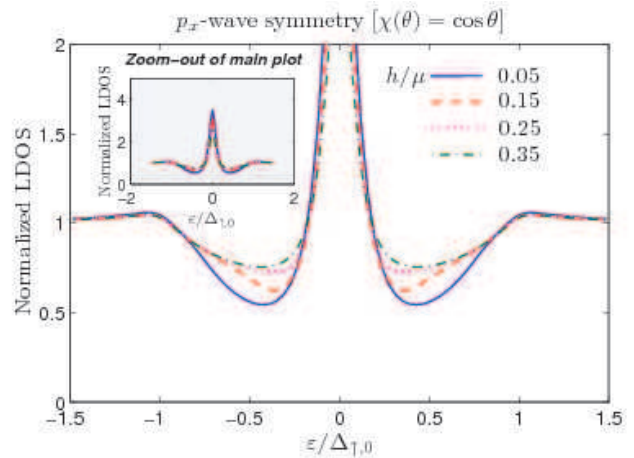}}
\caption{(Color online) Normalized LDOS in the vortex core for a $p_x$-wave symmetry [$\chi(\theta)=\cos\theta$] using 
several values of $h/\mu$.} 
\label{fig:Fig1}
\end{figure}

\begin{figure}[h!]
\centering
\resizebox{0.48\textwidth}{!}{
\includegraphics{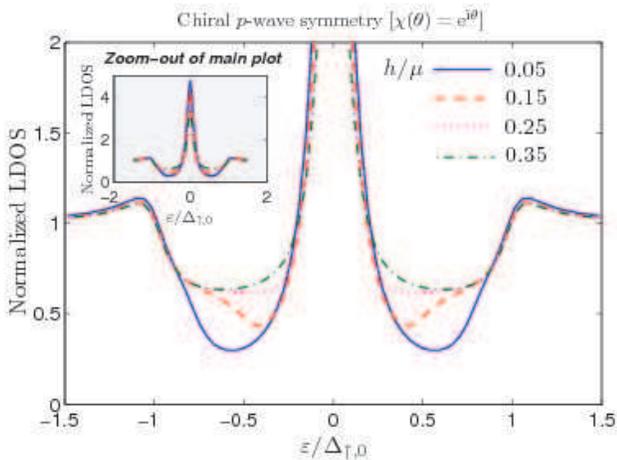}}
\caption{(Color online) Normalized LDOS in the vortex core for a chiral $p$-wave symmetry [$\chi(\theta)=\e{\i\theta}$] using 
several values of $h/\mu$. }
\label{fig:Fig2}
\end{figure}
\begin{figure}[h!]
\centering
\resizebox{0.48\textwidth}{!}{
\includegraphics{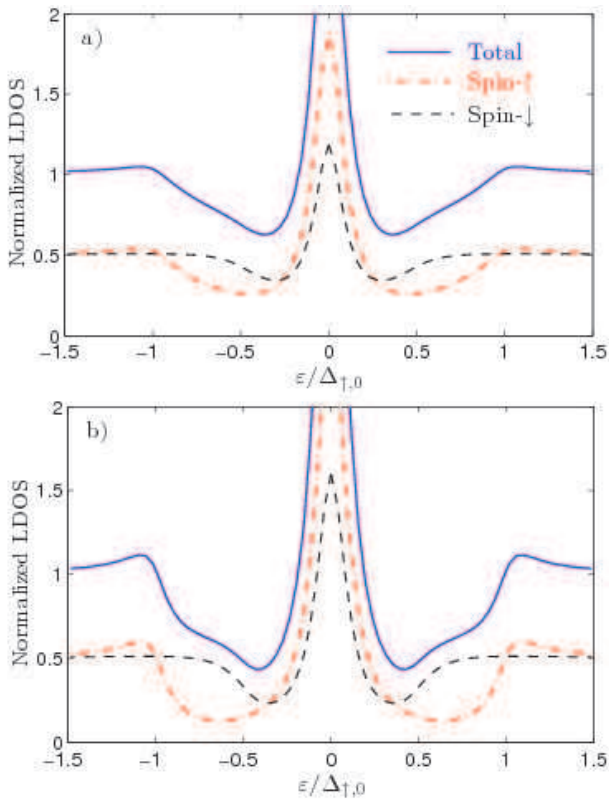}}
\caption{(Color online) Total and spin-resolved LDOS in the vortex core for the a) $p_x$-wave symmetry symmetry [$\chi(\theta)=\cos\theta$] and the b) chiral $p$-wave symmetry [$\chi(\theta)=\e{\i\theta}$]. We have here used $h/\mu=0.15$.}
\label{fig:spindos}
\end{figure}

\section{Results}\label{sec:results}

We begin by plotting the energy-resolved LDOS in the vortex core ($r_a=r_b=0$) for an order parameter which has line nodes 
in momentum space. Such an order parameter was recently proposed to be realized in UGe$_2$ by Harada \etal \cite{harada}, 
and it was moreover argued that the superconducting pairing only took place in the majority spin-band. To investigate 
how the relative magnitudes of the majority and minority spin gaps affect the LDOS in the vortex-core, we plot the LDOS 
for several values of the ratio $h/\mu$ in Fig. \ref{fig:Fig1}. As usual, the LDOS is strongly enhanced for subgap 
values due to the existence of bound states within the vortex core \cite{caroli}. The presence of two gaps in the 
system should manifest itself in the form of non-monotonous behaviour in the subgap spectrum, but it is not 
possible to discern such behaviour unambiguously from Fig. \ref{fig:Fig1}. This effect may be masked by strong 
inelastic scattering, modelled here by the parameter $\delta$, which effectively smears the LDOS. The effect of 
increasing the exchange field is seen to suppress the deviation from the normal-state LDOS. This may be 
understood by noting that the minority-spin gap is strongly reduced with increasing exchange field, and
that the corresponding increase of the majority-spin gap is not able to compensate for the suppressed 
regime of bound states within the core.
\par
We next study the chiral $p$-wave symmetry analogous to the A2-phase in liquid $^3$He, and plot the energy-resolved LDOS for 
several values of $h/\mu$ in Fig. \ref{fig:Fig2}. Although the qualitative behaviour is quite similar to Fig. \ref{fig:Fig1}, 
there are two important distinctions. First, one notices that the chiral symmetry appears to have a much more pronounced 
influence on the LDOS quantitatively, yielding a larger zero energy-peak and larger subgap dips. This is in fact \textit{opposite} 
what one would have expected from tunneling conductance measurements of $p_x$-wave and chiral $p$-wave superconductors, 
respectively. For such measurements, the zero-energy peak becomes much larger in the $p_x$-wave case than in the chiral 
$p$-wave case. Secondly, the subgap features associated with the presence of two gaps are enhanced in Fig. \ref{fig:Fig2} 
compared to Fig. \ref{fig:Fig1}. The non-monotonous behaviour for subgap energies is present for all curves in Fig. \ref{fig:Fig2}, but the features indicative of multiple gaps are most clearly seen for 
$h/\mu=0.15$, manifested through an additional inflection point before the normal state LDOS is recovered. These differences could be helpful in discriminating between different types 
of pairing symmetries in ferromagnetic superconductors.

In order to show more clearly the contribution from each spin-band to the LDOS near the vortex core, consider Fig. \ref{fig:spindos} 
where we plot the total LDOS and the contribution from each spin band for a) $\chi(\theta)=\cos\theta$ and b) $\chi(\theta)=\e{\i\theta}$. 
The rise of the LDOS following the gap edge $\Delta_{\sigma,0}$ of each spin band occurs at different energies due to the exchange 
splitting. This is revealed in the total LDOS as kinks located at two distinct energies, which offers the opportunity to obtain 
explicit information about the relative magnitude of the two gaps. The qualitative features are the same in 
Fig. \ref{fig:spindos} a) and b), but they are quantitatively more pronounced in the chiral $p$-wave symmetry case. This may be 
due to the fact that the chiral $p$-wave gap has a constant magnitude ($|\chi(\theta)|=1$), while the $p_x$-wave gap varies 
in magnitude upon traversing around the Fermi surface. Therefore, the LDOS is more strongly affected in the chiral $p$-wave case. 
\par
We now study the resolution of the LDOS in real space for a fixed energy in Fig. \ref{fig:contour1}. We have chosen $R=2$, 
corresponding to $h/\mu \simeq 0.14$ and also chosen the line node symmetry $\chi(\theta) = \cos\theta$. In all cases, the 
plots in Fig. \ref{fig:contour1} display a two-fold spatial symmetry, in accordance with the superconducting order parameter. \cite{schopohl,Hess,Ichioka} 
The zero-energy peak present for $\varepsilon=0$ evolves into a dip-structure at the vortex core upon increasing the 
quasiparticle energy. The deviation from the normal-state LDOS is still significant even at distances $\sim 2\xi$ away 
from the vortex core around $\varepsilon/\Delta_{\uparrow,0}=0.5$. The qualitative features are the same for the 
chiral $p$-wave symmetry in Fig. \ref{fig:contour2}, although the symmetry is now circular due to the isotropy 
of the magnitude of the gap ($|\chi(\theta)|=1$)

\begin{widetext}
\text{ }\\
\begin{figure}[h!]
\centering
\resizebox{1.0\textwidth}{!}{
\includegraphics{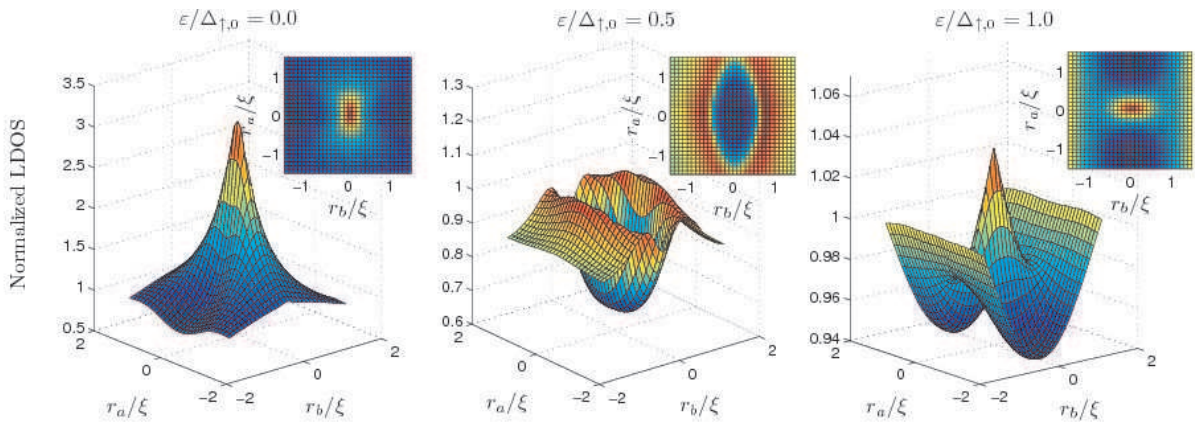}}
\caption{(Color online) Normalized LDOS in the vicinity of the vortex core at three different quasiparticle energies, using 
$R=2$ with a $p_x$-wave symmetry [$\chi(\theta)=\cos\theta$]. A two-fold symmetry is observed in agreement with the symmetry 
of the order parameter. }
\label{fig:contour1}
\end{figure}

\text{ }\\

\begin{figure}[h!]
\centering
\resizebox{1.0\textwidth}{!}{
\includegraphics{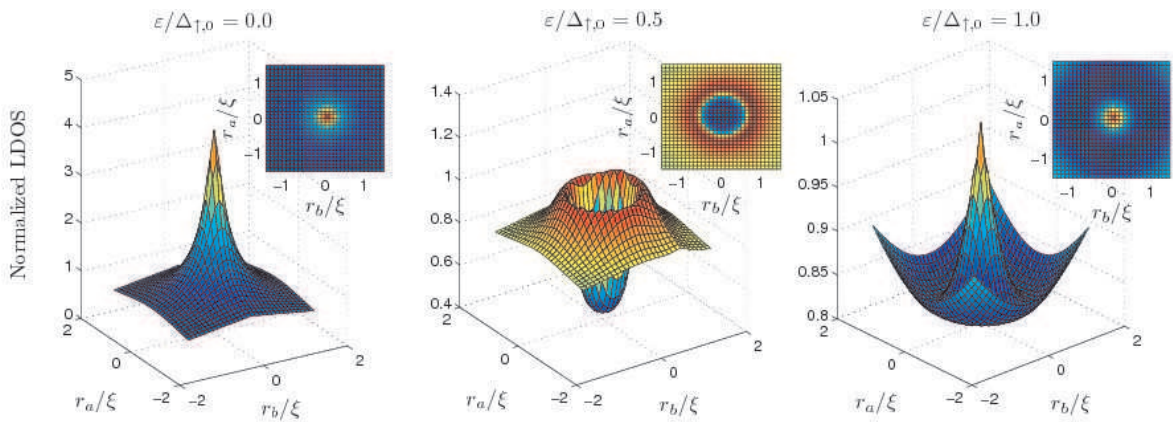}}
\caption{(Color online) Normalized LDOS in the vicinity of the vortex core at three different quasiparticle energies, using $R=2$ 
with a chiral $p$-wave symmetry [$\chi(\theta)=\e{\i\theta}$]. A circular symmetry is observed in agreement with the symmetry 
of the order parameter. }
\label{fig:contour2}
\end{figure}
\end{widetext}

\section{Discussion}\label{sec:discuss}

In our calculations, we have chosen a real gauge for both superconducting order parameters $\Delta_{\sigma}$, 
$\sigma=\uparrow,\downarrow$. If the two spin-bands are completely independent, there is no phase-locking between 
the order parameters which fixes the relative phase $\Delta\nu = \nu_\uparrow-\nu_\downarrow$, where $\nu_\sigma$ 
is phase associated with the broken U(1) symmetry. The existence of two such phases would imply that a U(1)$\times$U(1) 
symmetry is broken in a ferromagnetic superconductor, and would in principle allow for two critical temperatures 
which may differ in magnitude. However, if the two spin-bands do communicate by means of \eg spin-orbit coupling 
or impurity scattering, a term of the form $-\lambda\cos(\Delta\nu)$ will appear in the free energy describing the 
system. This corresponds to a phase-locking scenario where the sign of $\lambda$ determines whether $\Delta\nu = 0$ 
or $\Delta\nu = \pi$ is the energetically preferred relative phase. Above, we have decoupled the two spin-bands 
such that the relative phase of $\Delta_\uparrow$ and $\Delta_\downarrow$ is of no consequence. Taking into 
account scattering between the spin-bands would require solving coupled Ricatti-equations and investigating 
the effect of phase-locking explicitly, which is beyond the scope of this paper. 
\par
In Fig. \ref{fig:spindos}, we plotted the relative contribution from the two spin-bands to the LDOS near the vortex 
core to clarify how the LDOS may give decisive clues about whether both spin bands partake in the superconducting 
pairing or not. In principle, it might be possible to probe explicitly the spin-resolved LDOS by using a strong 
ferromagnetic STM and contrasting parallel and antiparallel relative configuration of the exchange fields in the 
FMSC and the ferromagnetic STM tip. The experimental realization of this particular proposal is nevertheless probably 
challenging.

\section{Summary}\label{sec:summary}

In summary, we have numerically studied the local density of states (LDOS) in the vicinity of a vortex core in a ferromagnetic 
superconductor. Specifically, we have investigated what influence the exchange field and the symmetry of the 
superconducting order parameter exhibit on both the spatially resolved and energy-resolved LDOS. The symmetry 
of the spatially resolved LDOS near the vortex core as revealed by STM-measurements should give decisive clues 
about the orbital symmetry of the superconducting order parameter,\cite{schopohl,Hess,Ichioka}  while the 
energy-resolved LDOS could provide important information about the presence of multiple gaps in the system. 
Our results should be comparable to experimentally obtained data, both qualitatively and quantitatively, and 
may thus be helpful in clarifying the nature of the superconducting pairing in ferromagnetic superconductors.

\acknowledgments

J.L. acknowledges A. Nevidomskyy for useful discussions. J.L. and A.S. were supported by the Norwegian Research Council Grant Nos. 158518/431, 158547/431, (NANOMAT), 
and 167498/V30 (STORFORSK). T.Y. acknowledges support by the JSPS.

\appendix 

\section{Matrices and quasiclassical theory}\label{app:eilenberger}
The matrices used in this paper are defined as \cite{jpdiplom}
\begin{align}
\underline{\tau_1} &= \begin{pmatrix}
0 & 1\\
1 & 0\\
\end{pmatrix},\;
\underline{\tau_2} = \begin{pmatrix}
0 & -\i\\
\i & 0\\
\end{pmatrix},\;
\underline{\tau_3} = \begin{pmatrix}
1& 0\\
0& -1\\
\end{pmatrix},\notag\\
\underline{1} &= \begin{pmatrix}
1 & 0\\
0 & 1\\
\end{pmatrix},\;
\hat{1} = \begin{pmatrix}
\underline{1} & \underline{0} \\
\underline{0} & \underline{1} \\
\end{pmatrix},\;
\hat{\tau}_i = \begin{pmatrix}
\underline{\tau_i} & \underline{0}\\
\underline{0} & \underline{\tau_i} \\
\end{pmatrix},\notag\\
\hat{\rho}_1 &= \begin{pmatrix}
\underline{0} & \underline{\tau_1}\\
\underline{\tau_1} & \underline{0} \\
\end{pmatrix},\;
\hat{\rho}_2 =  \begin{pmatrix}
\underline{0} & -\i\underline{\tau_1}\\
\i\underline{\tau_1} & \underline{0} \\
\end{pmatrix},\;
\hat{\rho}_3 = \begin{pmatrix}
\underline{1} & \underline{0}\\
\underline{0} & -\underline{1}  \\
\end{pmatrix}.
\end{align}

Let us briefly sketch the way to obtian the quasiclassical Eilenberger equations for a non-unitary, spin-triplet superconducting state coexisting with ferromagnetism. For further background information on the quasiclassical theory of superconductivity, the reader may consult \eg Refs.~\onlinecite{serene, kopnin, rammer, zagoskin, Chandrasekhar} for nice reviews. We follow here closely the notation of Ref. \cite{jpdiplom}. Our starting point is the following Hamiltonian:
\begin{align}
H &= \sum_{\alpha\beta} \int \text{d}\vecr \psi^\dag_\alpha(\vecr,t)\Big(-\frac{\nabla^2}{2m}\underline{1} - h\underline{\tau_3} \Big)_{\alpha\beta} \psi_\beta(\vecr,t) \notag\\
&- \sum_\sigma \int \text{d}\vecr \text{d}\vecr' [\Delta_\sigma(\vecr,\vecr') \psi^\dag_\sigma(\vecr)\psi^\dag_\sigma(\vecr) \notag\\
&+ \Delta_\sigma^*(\vecr,\vecr')\psi_\sigma(\vecr')\psi_\sigma(\vecr)].
\end{align}
The Heisenberg equation of motion for the above Hamiltonian is obtained in the standard way:
\begin{align}\label{eq:heis}
\i \partial_t \hat{\rho}_3 \Psi(\vecr,t) &= \int \text{d}\vecr' \hat{H}(\vecr,\vecr',t) \Psi(\vecr,t),\notag\\
\hat{H}(\vecr,\vecr',t) &= \hat{\xi}(\vecr)\delta(\vecr-\vecr') - \hat{\Delta}(\vecr,\vecr'),\; \hat{\xi}(\vecr) = -\frac{\nabla_\vecr^2}{2m} \hat{1},\notag\\
\hat{\Delta}(\vecr,\vecr') &= \begin{pmatrix} 
\underline{0} & \underline{\Delta}(\vecr,\vecr') \\
\underline{\Delta}^*(\vecr,\vecr')  & \underline{0} \\
\end{pmatrix},\notag\\
\underline{\Delta}(\vecr,\vecr') &= \text{diag}[\Delta_\uparrow(\vecr,\vecr'),\Delta_\downarrow(\vecr,\vecr')].
\end{align}
For simplicity, we consider only the retarded component of the Green's function $G^\text{R}$ in what follows, since the system is specified exclusively by $G^\text{R}$ in an equilibrium situation. It is defined as
\begin{align}
\underline{G}_{\alpha\beta}^\text{R}(1,2) &= -\i\Theta(t_1-t_2)\langle [\psi_\alpha(1),\psi_\beta^\dag(2)]_+ \rangle,
\end{align}
where the notation $(1,2)$ refers to the spatial and time coordinates: $(1) \equiv (\mathbf{r}_1;t_1)$. We explicitly write the $'+'$ sign as a subscript to denote an anticommutator; it is else implicitly understood that the notation $[...]$ denotes a usual commutator. Similarly, the anomalous Green's function is given by
\begin{align}
\underline{F}_{\alpha\beta}^\text{R}(1,2) &= -\i\Theta(t_1-t_2)\langle [\psi_\alpha(1),\psi_\beta(2)]_+ \rangle.
\end{align}
One may construct $4\times4$ matrices in combined particle-hole and spin space, known as Nambu space, in the following manner:
\begin{align}
\hat{G}^\text{R}(1,2) = \begin{pmatrix}
\underline{G}^\text{R}(1,2) & \underline{F}^\text{R}(1,2)\\
[\underline{F}^\text{R}(1,2)]^* & [\underline{G}^\text{R}(1,2)]^*\\
\end{pmatrix}.
\end{align}
Note that $G(1,2)$ is a generalized Gor'kov Green's function, which contains information about processes occuring at length scales comparable to the Fermi wavelength. Such information is lost upon applying the quasiclassical approximation. Using the Heisenberg equation of motion Eq. (\ref{eq:heis}), we obtain
\begin{widetext}
\begin{align}\label{eq:eqofmot}
\Bigg[ \i\partial_{t_1}\Big( \hat{\rho}_3 \hat{G}^\text{R}(1,2)\Big)_{ij} - \int \text{d}\vecr' \sum_l [-\i\Theta(t_1-t_2)] \hat{H}_{il} (\vecr_1,\vecr',t_1) (\hat{\rho}_3)_{ll} \langle [\Psi_l(\vecr',t_1),\Psi_j^\dag(\vecr_2,t_2)]_+ \rangle \Bigg] = \delta_{ij}\delta(1-2).
\end{align}
\end{widetext}
To arrive at Eq. (\ref{eq:eilenberger}), it is convenient to introduce the mixed representation which shifts the frame of reference to a center-of-mass system. We define
\begin{align}
\mathbf{R} &= (\mathbf{r}_1+\mathbf{r}_2)/2,\; \mathbf{r} = \mathbf{r}_1 - \mathbf{r}_2,\notag\\
T &= (t_1+t_2)/2,\; t = t_1-t_2,
\end{align}
such that
\begin{equation}\label{eq:b1}
\hat{G}^\text{R}(1,2) = \hat{G}^\text{R}(\mathbf{R}+\frac{\mathbf{r}}{2},T+\frac{t}{2},\mathbf{R} - \frac{\mathbf{r}}{2},T-\frac{t}{2}).
\end{equation}
The Fourier-transformation of Eq. (\ref{eq:b1}) yields
\begin{align}\label{eq:b2}
\hat{G}^\text{R}(\mathbf{p},\mathbf{R};T,\varepsilon) = \int \text{d}\mathbf{r}\e{-\i\vp\mathbf{r}} \int \text{d}t \e{\i t\varepsilon} \hat{G}^\text{R}(1,2).
\end{align}
An exact solution for $\hat{G}^\text{R}(\mathbf{p},\mathbf{R};T,\varepsilon)$ is very hard to achieve, but the situation is considerably 
simplified if one is willing to neglect all atomic-scale fine structure effects that are included in $\hat{G}^\text{R}$. These give rise 
to a rapidly oscillating part in the solution for $\hat{G}^\text{R}$, and rewriting the Green's function through Eq. (\ref{eq:b2}) allows 
us to integrate out this unnecessary information (at least for our purposes). This approximation may be expected to yield 
satisfactory results if the energy of the physical quantities involved in the problem, \eg exchange field and superconducting 
order parameter, are much smaller than the Fermi energy. Assuming that only particles in the vicinity of Fermi level will take 
part in physical processes, one only needs to retain the direction of the momentum at Fermi level in the $\mathbf{p}$ coordinate. 
\par
As this Appendix is only meant as background information for the Eilenberger equation, we do not show all the details leading 
from Eq. (\ref{eq:eqofmot}) to Eq. (\ref{eq:eilenberger}) here. The calculations are nevertheless fairly straight-forward, and 
consist of first switching to a mixed representation, then Fourier-transforming the variables, and finally performing the 
quasiclassical approximation
\begin{align}
\hat{g}^\text{R} = \frac{\i}{\pi} \int \text{d}\xi_\vp \hat{G}^\text{R},\; \xi_\vp = \frac{\vp^2}{2m}.
\end{align}

\section{Inelastic scattering}\label{app:inelastic}
\begin{figure}[h!]
\centering
\resizebox{0.45\textwidth}{!}{
\includegraphics{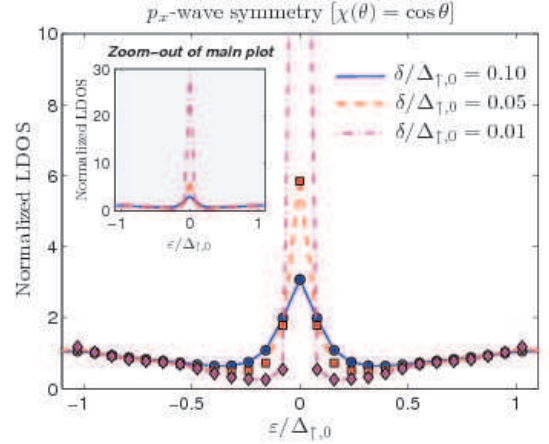}}
\caption{(Color online) Normalized LDOS in the vicinity of the vortex core at three different values of the inelastic scattering rate $\delta=\tau^{-1}$, using a $p_x$-wave symmetry with $h/\mu=0.15$. }
\label{fig:inelastic}
\end{figure}

The choice of $\delta=0.1\Delta_{\uparrow,0}$ is motivated by the fact that the zero-energy peaks observed in experiments are usually limited from above to roughly a factor of five times the normal-state value of the LDOS, which we reproduce with this particular choice of $\delta$. Choosing $\delta$ smaller (corresponding to a longer quasiparticle lifetime since $\delta=\tau^{-1}$) causes the zero-energy peak to grow substantially, as shown in Fig. \ref{fig:inelastic}. In general, the inelastic scattering rate does not have to be proportional to the gap at all and our choice of $\delta=0.1\Delta_{\uparrow,0}$ is simply chosen to compare the scattering rate against a familiar quantity.

\section{Derivation of Eq. (\ref{eq:gap})}\label{app:gap}

The gap equation may be obtained by starting out with a Hamiltonian assuming a non-unitary triplet pairing state coexisting with itinerant ferromagnetism \cite{nevidomskyy, gronsleth}, namely
\begin{align}
\hat{H} &= \sum_\vk \xi_\vk + \frac{INM^2}{2} - \frac{1}{2}\sum_{\vk\sigma} \Delta_{\vk\sigma\sigma}^\dag b_{\vk\sigma\sigma} \notag\\
&+\frac{1}{2}\sum_{\vk\sigma} \Big(\hat{c}_{\vk\sigma}^\dag \hat{c}_{-\vk\sigma}\Big)
\begin{pmatrix}
\xi_{\vk\sigma} & \Delta_{\vk\sigma\sigma} \\
\Delta_{\vk\sigma\sigma}^\dag & -\xi_{\vk\sigma} \\
\end{pmatrix}
\begin{pmatrix}
\cop_{\vk\sigma}\\
\cop_{-\vk\sigma}^\dag\\
\end{pmatrix},
\end{align}
Here, $I$ is the ferromagnetic exchange coupling constant, $N$ is the number of lattice sites, $M$ denotes the magnetic order parameter (dimensionless), while $b_{\vk\sigma\sigma}$ is the Cooper pair expectation value. Diagonalization of this Hamiltonian produces:
\begin{align}
\hat{H} &= H_0 + \sum_{\vk\sigma} E_{\vk\sigma} \hat{\gamma}_{\vk\sigma}^\dag\hat{\gamma}_{\vk\sigma},\notag\\
H_0 &= \frac{1}{2}\sum_{\vk\sigma}(\xi_{\vk\sigma} - E_{\vk\sigma} - \Delta_{\vk\sigma\sigma}^\dag b_{\vk\sigma\sigma}) + \frac{INM^2}{2},
\end{align}
where $\{\hat{\gamma}_{\vk\sigma},\hat{\gamma}_{\vk\sigma}^\dag\}$ are new fermion operators and the eigenvalues read
\begin{equation}
E_{\vk\sigma} = \sqrt{\xi_{\vk\sigma}^2 + |\Delta_{\vk\sigma\sigma}|^2}.
\end{equation}
Above, $\xi_\vk$ is the kinetic energy measured from Fermi level. By minimizing the free energy, one obtains the gap equation for the superconducting order parameter: \cite{nevidomskyy}
\begin{align}
\Delta_{\vk\sigma\sigma} = -\frac{1}{N}\sum_{\vk'} V_{\vk\vk'\sigma\sigma} \frac{\Delta_{\vk'\sigma\sigma}}{2 E_{\vk'\sigma}}\text{tanh}(\beta E_{\vk'\sigma}/2).
\end{align}
Assuming that the gap is fixed on the Fermi surface in the weak-coupling limit, one may write in general
\begin{equation}\label{eq:gaptriplet}
V_{\sigma\sigma}(\theta,\theta') = -V_0Y^\sigma(\theta)[Y^\sigma(\theta')]^*.
\end{equation}
where $Y^\sigma(\theta)$ are basis functions for the angular dependence of the interaction. To model $p_x$-wave and chiral $p$-wave pairing, respectively, we use $Y^\sigma(\theta)=-\sigma\e{\i\sigma\theta}$ and $Y^\sigma(\theta)=\cos\theta$. 
Conversion to integral gap equations is accomplished by means of the identity
\begin{equation}
\frac{1}{N} \sum_\vk f(\xi_{\vk\sigma}) = \int \text{d}\varepsilon N^\sigma(\varepsilon),
\end{equation}
where $N^\sigma(\varepsilon)$ is the spin-resolved density of states. In three spatial dimensions, this may be calculated from the dispersion relation by using the formula
\begin{equation}
N^\sigma(\varepsilon) = \frac{V}{(2\pi)^3} \int_{\varepsilon_{\vk\sigma} = \text{const}} \frac{\text{d} S_{\varepsilon_{\vk\sigma}}}{|\hat{\nabla}_\vk \varepsilon_{\vk\sigma}|}.
\end{equation}
With the dispersion relation $\xi_{\vk\sigma}= \varepsilon_\vk - \sigma IM - \mu$, one obtains
\begin{equation}
N^\sigma(\varepsilon) = \frac{mV\sqrt{2m(\varepsilon + \sigma IM + \mu)}}{2\pi^2}.
\end{equation}
In their integral form, the gap equation reads
\begin{align}\label{eq:gapeqint}
1 &= \frac{V_0}{4\pi}\sum_\sigma \int^{\omega_0}_{-\omega_0} \text{d}\varepsilon \frac{ N^\sigma(\varepsilon) Y^\sigma(\theta)[Y^\sigma(\theta')]^* }{E_\sigma(\varepsilon)}\text{tanh}[\beta E_\sigma(\varepsilon)/2].
\end{align}
Consider now $T=0$, where the integral may be done analytically to yield:
\begin{equation}\label{eq:gapanalytical}
\Delta_{\sigma,0} = c\omega_0 \e{-1/g\sqrt{1+\sigma\tilde{M}}},\; \sigma=\uparrow,\downarrow
\end{equation}
where we have defined $\tilde{M} = IM/\mu=h/\mu$, \ie the exchange energy scaled on the Fermi energy. Moreover, $c$ is a numerical prefactor which depends on which symmetry one considers ($p_x$-wave or chiral $p$-wave) while $g$ is the weak-coupling constant. The important influence of the magnetization is that it modifies the density of states, which affects the superconductivity gaps. For $\tilde{M} = 1$, i.e. an exchange splitting equal to the Fermi energy, the minority spin gap is completely suppressed. Thus, the presence of magnetization reduces the available phase space for the minority spin Cooper pairs, suppressing the gap and the critical temperature compared to the pure Bardeen-Cooper-Schrieffer case.

\end{document}